\documentclass[sigconf]{acmart}

\usepackage{multirow}
\usepackage{multicol}
\usepackage{tabularx}
\usepackage{xspace}
\usepackage{bm}
\usepackage[linesnumbered,ruled]{algorithm2e}
\usepackage{algorithmic}
\usepackage[misc]{ifsym}

\usepackage{bbding}

\newcommand{\paratitle}[1]{\vspace{1.5ex}\noindent\textbf{#1}}
\newcommand{\eg}{\emph{e.g.,}\xspace}
\newcommand{\ie}{\emph{i.e.,}\xspace}

\newcommand{\wrt}{w.r.t.\xspace}
\newcommand{\ignore}[1]{}
\newcommand{\ourmodel}{UniM$^2$Rec}

\AtBeginDocument{%
  \providecommand\BibTeX{{%
    \normalfont B\kern-0.5em{\scshape i\kern-0.25em b}\kern-0.8em\TeX}}}

\setcopyright{acmcopyright}
\copyrightyear{2023}
\acmYear{2023}
\acmDOI{XXXXXXX.XXXXXXX}

\acmConference[Conference acronym 'XX]{Make sure to enter the correct
  conference title from your rights confirmation emai}{June 03--05,
  2023}{Woodstock, NY}
\acmPrice{15.00}
\acmISBN{978-1-4503-XXXX-X/23/06}

\begin{document}

\title{Universal Multi-modal Multi-domain Pre-trained Recommendation}

\author{Wenqi Sun\textsuperscript{\textmd{1}},
Ruobing Xie\textsuperscript{\textmd{2}},
Shuqing Bian\textsuperscript{\textmd{3}},
Wayne Xin Zhao\textsuperscript{\textmd{1}},
Jie Zhou\textsuperscript{\textmd{2}}}

\email{wenqisun@ruc.edu.cn,xrbsnowing@163.com,batmanfly@gmail.com}

    \affiliation{%
  \institution{\textsuperscript{\textmd{1}}Gaoling School of Artificial Intelligence, Renmin University of China}
  \institution{\textsuperscript{\textmd{2}}Tencent Inc.}
  \institution{\textsuperscript{\textmd{3}}School of Information, Renmin University of China}
   \country{}
   }

\renewcommand{\authors}{Wenqi Sun, Ruobing Xie, Shuqing Bian, Wayne Xin Zhao, Jie Zhou}
\renewcommand{\shortauthors}{Wenqi Sun, Ruobing Xie, Shuqing Bian, Wayne Xin Zhao, Jie Zhou}

\begin{abstract}
There is a rapidly-growing research interest in modeling user preferences via pre-training multi-domain interactions for recommender systems.
However, Existing pre-trained multi-domain recommendations mostly select the item texts to be bridges across domains, and simply explore the user behaviors in target domains. Hence, they ignore other informative multi-modal item contents (\eg visual information), and also lack of thorough consideration of user behaviors from all interactive domains.
To address these issues, in this paper, we propose to pre-train \underline{Uni}versal \underline{M}ulti-modal item content presentation for \underline{M}ulti-domain \underline{Rec}ommendation, called \textbf{\ourmodel},
which could smoothly learn the multi-modal item content presentations and the multi-modal user preferences from all domains.
With the pre-trained multi-domain recommendation model, \ourmodel~could be efficiently and effectively transferred to new target domains in practice.
Extensive experiments conducted on five real-world datasets in target domains demonstrate the superiority of the proposed method over existing competitive methods, especially for the real-world recommendation scenarios that usually struggle with seriously missing or noisy item contents.
\end{abstract}



\maketitle

\section{Introduction}

As a leading approach, pre-training has yielded significant breakthroughs for its strong ability of learning from a large volume of data for multiple downstream tasks in the fields of NLP and CV~\cite{GPT-2,T5,kosmos-1,llm2023}. But in some areas where the knowledge is personalized and not general under most conditions, such as recommender systems (RS), the application of pre-training still faces various challenges.
In RS, the major downstream tasks are to recommend in different domains, where a user may have different preferences (\eg the same user may love modern science fiction movies and classical music). Therefore, it is a highly essential for pre-trained recommendation models to understand user preferences on \emph{different domains}.

\begin{figure}[!hbpt]
    \centering
    \includegraphics[width=0.99\linewidth]{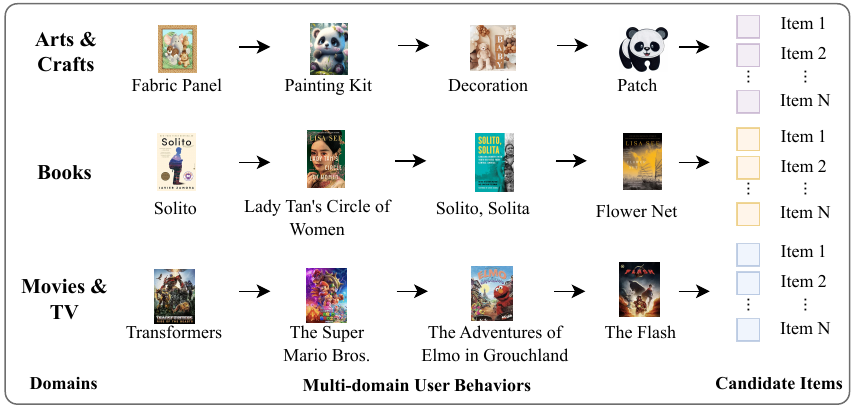}
    \caption{An illustration of multi-modal item contents and multi-domain user behaviors.}
    \label{fig:illustration}
\end{figure}

To address this domain inconsistency issue, multi-domain recommendation (MDR), which aims to improve user preference modeling in target domains using knowledge transferred from multiple source domains, has been studied~\cite{CCDR2022}.
The key challenge of multi-domain recommendation is how to conduct the informative positive transfer from multiple source domains to target domains. Typical MDR models often rely on overlapped users or items \cite{man2017cross,CCDR2022}.
Recently, with the thriving of pre-training models, some efforts attempt to introduce pre-trained item texts as anchors to store general knowledge in MDR. Specifically, UniSRec~\cite{UniSRec2022} utilizes item texts processed by pre-trained language model to get item representations as anchors in all domains.
Moreover, the discussions about ID- or modality-based recommendation models have attracted widespread attentions~\cite{IDMoRec2023}.
Hence, it is promising to build a universal and robust pre-trained recommendation model incorporating multi-modal item contents and multi-domain interactions.

Despite the progress, existing pre-training MDR models still struggle with two issues:
(1) Most MDR models mainly select the textual information to be the bridge across domains \cite{UniSRec2022}, ignoring other informative modality such as \emph{visual information}. Nowadays, visual information becomes increasingly dominant in practical recommendations due to the thriving of micro-video and photo sharing applications. To some extent, the visual modality constitutes the primary information source of items in lots of domains (\eg \emph{arts} and \emph{videos})~\cite{kosmos-1,kosmos-2}. Moreover, pre-trained MDR models should be robust with different downstream datasets, while real-world datasets usually struggle with missing or noisy item contents in some modalities. Simply relying on one modality as the bridge of MDR will weaken robustness of item content representation and multi-domain transferring.
(2) Most of these methods also lack of thorough consideration of user preferences from all interactive domains. 
Practical huge platforms (\eg YouTube, Instagram and WeChat) usually have multiple available domains with the shared user account systems, and thus it is beneficial to consider the user’s historical behaviors from all interactive domains in pre-trained MDR models.
Figure~\ref{fig:illustration} shows the multi-modal item contents from multiple domains, where there may be correlations among different interactive domains for the same user.
In particular, the visual modality is more dominant than textual modality on ``Arts and Crafts'', and it is the opposite effect on ``Books''. Both visual and textual modalities are important on ``Movies and TV''.
Hence, it is natural and indispensable to jointly consider item multi-modal contents and multi-domain interactions in pre-trained MDR, which will help to build a effective, universal and robust pre-trained MDR model.

In this work, we aim to design a universal pre-trained MDR framework that can learn the multi-modal item content representations and the user preferences from all interacted domains for better serving real-world multi-domain recommendation scenarios.
For this purpose, we propose \textbf{\ourmodel}, which could smoothly deal with the inputs of multi-modal item contents and the multi-domain user interactions.
Specifically, \ourmodel~mainly contains three modules, namely the multi-modal item content constructors, the multi-domain item projectors, and the mixed user behavior flow encoder.
Firstly, the multi-modal item content constructors are adopted to learn the multi-modal item content representations via the pre-trained models.
Next, \ourmodel~adopts the multi-domain item projectors, which consist of the parametric whitening and the mixture-of-experts (MoE) to project the domain-specific item contents into a universal item content representation space.
Finally, the universal item content representations of a user's historical behaviors in all interacted domains, previously projected into the shared item content representation space, are directly mixed in the chronological order and fed into the mixed user behavior flow encoder to learn the user preferences for next-item predictions in target domains.
With the pre-trained multi-domain recommendation model, \ourmodel~could be efficiently and effectively transferred to new target domains.
In experiments, we conduct extensive evaluations on multiple target domain datasets, where \ourmodel~achieves consistent improvements over competitive baseline methods. Moreover, we conduct the ablation study for in-depth understanding of each techniques in \ourmodel.
we also verify the robustness of \ourmodel~ on possible item content missing or noises and the few-shot item scenarios.
Our main contributions are summarized as follows:

\begin{itemize}
    \item We bring in the consideration of multi-modal item contents and user preferences from all interacted domains for the pre-trained MDR. To the best of our knowledge, \ourmodel~is the first pre-trained MDR model that introduce the item cross-modal representations to enhance the robustness of item content representations.
    \item We propose a novel universal and robust multi-modal item content presentation framework for multi-domain recommendation, which contains a set of effective modules including the multi-modal item content constructors, the multi-domain item projectors, and the mixed user behavior flow encoder.
    \item Extensive experiments conducted on five real-world datasets demonstrate the effectiveness and robustness of the proposed method. \ourmodel~is an indispensable step of the journey to practical pre-trained recommendations.
\end{itemize}

\section{Related Work}
\subsection{Sequential Recommendation}
The sequential recommendation task aims to understand and model the sequential behaviors of the users over time.
Early works (\eg FPMC~\cite{FPMC2010}) on sequential recommendation are primarily based on the Markov Chain assumption and focus on learning the item-item transitions to predict the next item.
With the development of deep learning techniques, RNN/CNN-based recommendation methods (\eg GRU4Rec~\cite{GRU4Rec}, NARM~\cite{NARM}, Caser~\cite{Caser}, and NextItNet~\cite{NextItNet}) were introduced to capture sequential patterns.
In recent years, Transformer-based methods have shown strong performance for sequential recommendation due to powerful sequential modeling capacity~\cite{SASRec,BERT4Rec,sun2023etrec}.
Further, CL4SRec \cite{CL4SRec2022} and S$^3$-Rec \cite{S3-Rec2020} enhance the sequential modeling with the techniques of various intra-domain contrastive learning and pre-training.
Despite the success of these sequential recommendation methods,
they often ignore the multi-domain user interactions and the multi-modal item contents.

\subsection{Multi-domain Recommendation}
Multi-domain recommendation (MDR) aims to improve the recommendation performance by transferring knowledge from multiple source domains to target domains~\cite{STAR2021}.
The key techniques \wrt multi-domain recommendation~\cite{AFT2021} and cross-domain recommendation (CDR) are similar~\cite{ADIN2022,Integrate2021}.
CDR mostly aims to transfer knowledge in a specific direction (\eg leveraging interactions from main scenarios to improve performance in cold-start scenarios)~\cite{CCDR2022,Feedback2020},
while MDR aims to gain an overall improvement in all interactive domains.
For example, $\pi$-Net~\cite{PINet2019} and PSJNet~\cite{PSJNet2023} employ RNN to generate user-specific representations,
which emphasize the sequential dependencies but fail to depict transitions among associated entities.
STAR~\cite{STAR2021} utilizes the star topology from shared centered parameters and domain-specific parameters for different domains.
AFT~\cite{AFT2021} leverages an adversarial learning method to solve the MDR problem.
CCDR~\cite{CCDR2022} proposes two intra-domain and inter-domain contrastive learning tasks to enhance cross-domain recommendation, where the aligned users, tags, words, and categories are functioned as the semantic bridges across different domains.
UniSRec~\cite{UniSRec2022} utilizes item texts processed by pre-trained language model to get the universal item representation as anchors in all domains.
In this work, we mainly focus on how to leverage the pre-training techniques to learn universal and robust multi-modal item content representations and multi-domain user behavior sequences.

\subsection{Multi-modal Recommendation}
Multi-modal recommendation exploits the multi-modal side information of items to improve recommendation performance.
Previous works incorporate the visual features of items into the matrix factorization based recommendation~\cite{VBPR2016,DVBPR2017,GraphCAR2018},
and attention networks combine multi-modal features to enhance the representation learning of users and items~\cite{MAML2019,PMGT2021,MRG2019}.
Recently, several studies~\cite{GRCN2020,LATTICE2021,sun2023generative,MVGAE2022,DualGNN2023,MARIO2022,BM3-2023,MMSSL2023,MMEKG2022} leverage Graph Neural Networks (GNNs) to exploit the multi-modal item information.
For example, LATTICE~\cite{LATTICE2021} embeds a modality-aware graph structure learning layer,
which identifies item-item graph structures by using multi-modal features.
DualGNN~\cite{DualGNN2023} explicitly models the user's attention over different modalities and inductively learns user preferences.
MVGAE~\cite{MVGAE2022} leverages a multi-modal variational graph auto-encoder to fuse modality-specific node embeddings according to the product-of-experts principle.
In another line of work,
MML~\cite{MML2022} incorporates the multi-modal side information to improve and stabilize the meta-learning process, aiming to alleviate the cold-start problem.
Therefore, it is highly essential for recommender systems to make full use of multi-modal item contents~\cite{Freedom2023,IDMoRec2023,AdaptiveModal2023,MISSRec2023}.

\section{Methodology}

\begin{figure*}[!t]
    \centering
    \includegraphics[width=0.95\linewidth]{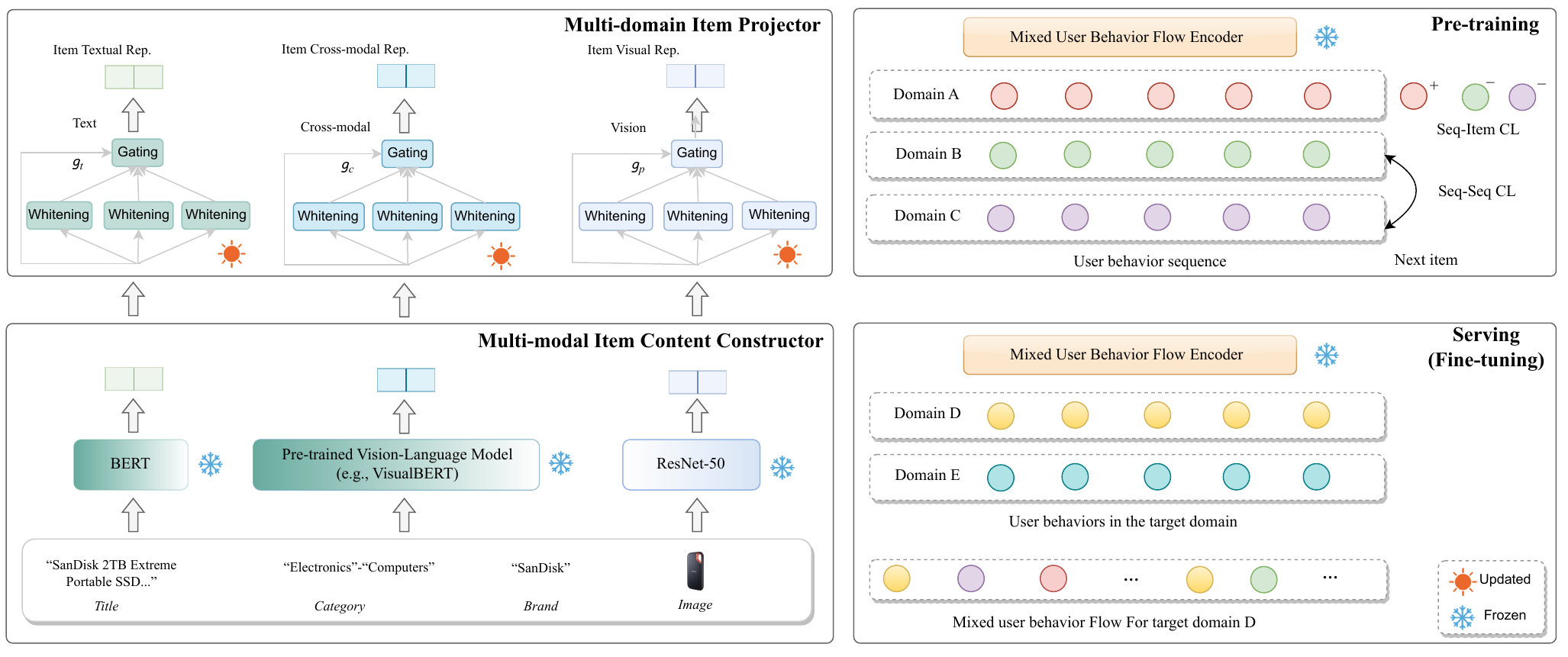}
    \caption{The overall framework of our proposed \ourmodel.}
    \label{fig:overview}
\end{figure*}

\subsection{Task Formulation}
\label{sec:overview}
The multi-domain recommendation task aims to improve the recommendation performance in target domains using knowledge transferred from the multiple source domains.
Let us take the source domains (\ie A and B) and the target domains (\ie C and D) as example, and formulate the MDR task as follows. 
We denote two source domain and two target domain behavior sequences of a user as $S_A =\{v^A_1,v^A_2,...,v^A_i,...\}$, $S_ B=\{v^B_1,v^B_2,...,v^B_j,...\}$, $S_ C=\{v^C_1,v^C_2,...,v^C_k,...\}$, and $S_D=\{v^D_1,v^D_2,...,v^D_\ell,...\}$, where $v^A_i$, $v^B_j$, $v^C_k$ and $v^D_\ell$ are the interacted items in domain $A$, $B$, $C$ and $D$, respectively. 
Then, the mixed source domain behavior sequence $S_M$ is produced by merging $S_A$ and $S_B$ in the chronological order, for which the above example can be given as $S_M = \{...,v^A_i,v^B_j,...\}$.
Given $S_A$, $S_B$, $S_M$ in source domains, and the interacted items in target domains $C$ and $D$, 
MDR tries to predict the next item that the user would interact with in target domains.
Note that our proposed method do not necessarily require overlapping users between source and target domains, instead, 
we aim to devise a universal and robust pre-trained MDR framework for learning multi-modal item content representations and user preferences from all interacted domains.
The overall framework of \ourmodel~is depicted in Figure~\ref{fig:overview}.

\subsection{Multi-modal Item Content Constructor}
\label{sec:itemrep}
Considering the strong modeling capacity of pre-trained models (\eg BERT~\cite{BERT2019}, ResNet~\cite{ResNet2016} and VisualBERT~\cite{visualbert}), we utilize them them to construct the semantic representations of items.
To be specific, given an item $v_i$ and its corresponding text $t_i$ (\ie title, category and brand) and image $p_i$,
we utilize BERT and ResNet-$50$ to learn the item textual and visual representations.
In addition, we leverage VisualBERT~\cite{visualbert}, a widely used pre-trained vision-language model, to obtain the cross-modal item representation so as to enhance the robustness of item content representations,
because of the excellent multi-modal modeling capacity of pre-trained vision-language models.
Note that these pre-trained models (\ie BERT, ResNet-50 and VisualBERT) are frozen and not to be fine-tuned.
After that,
the text feature of $t_i$ are extracted by BERT to form a hidden vector $\mathbf{e}_{t_i}\in \mathbb{R}^{d_1}$.
The image feature of $p_i$ are extracted by ResNet-$50$ to form a hidden vector $\mathbf{e}_{p_i}\in \mathbb{R}^{d_2}$.
The cross-modal fusion feature of $t_i$ and $p_i$ are extracted by VisualBERT to form a hidden vector $\mathbf{e}_{c_i}\in \mathbb{R}^{d_3}$, where $d_1$, $d_2$, $d_3$ are the dimensions of the hidden vectors.
To sum up, we obtain the textual, visual and cross-modal semantic representations of item $v_i$ (\ie $\mathbf{e}_{t_i}$, $\mathbf{e}_{p_i}$ and $\mathbf{e}_{c_i}$).

It is worth noting that the cross-modal item content representations from pre-trained vision-language models are highly essential, in that some items may have text missing, while others may have image missing in practical recommender systems, therefore, 
there are missing or noise in item textual or visual representations  when modeling item texts and images separately.
In contrast, as long as there exist texts or images for the items, the cross-modal item content representations are robust.
Therefore, the cross-modal item content representations are useful to enhance the robustness in real-world recommendation scenarios.

\subsection{Multi-domain Item Projector}
\label{sec:projector}

Though we obtain the semantic representations of items,
they do not directly fit into the recommendation task very well, because of large semantic gaps between the user behavior modeling in RS and the semantic representations of items in NLP/CV~\cite{SemanticGap2020}.
Inspired by recent works on the whitening-based methods~\cite{Whiten},
we conduct a linear transformation by whitening transformation to alleviate redundant correlations to some extent and derive isotropic semantic representations of items. 
Specifically, we design the parametric whitenings separately for texts, images and cross-modalities and incorporate learnable parameters in the parametric whitening transformations, respectively.
Formally, we have:
\begin{align}
    \label{eq:whitening_1}
   \widetilde{\mathbf{e}}_{t_i} = &(\mathbf{e}_{t_i}-\bm{b}_1) \cdot \bm{W}_1,\\
   \label{eq:whitening_2}
   \widetilde{\mathbf{e}}_{p_i}=&(\mathbf{e}_{p_i}-\bm{b}_2) \cdot \bm{W}_2,\\ 
    \label{eq:whitening_3}
    \widetilde{\mathbf{e}}_{c_i} =&(\mathbf{e}_{c_i}-\bm{b}_3) \cdot \bm{W}_3, 
\end{align}
where $\bm{b}_1\in \mathbb{R}^{d_1}$, $\bm{b}_2\in \mathbb{R}^{d_2}$, $\bm{b}_3\in \mathbb{R}^{d_3}$
and $\bm{W}_1\in \mathbb{R}^{d_1 \times d}$, $\bm{W}_2\in \mathbb{R}^{d_2 \times d}$, $\bm{W}_3\in \mathbb{R}^{d_3 \times d}$ are learnable parameters,
and $\widetilde{\mathbf{e}}_{t_i}$, $\widetilde{\mathbf{e}}_{p_i}$, $\widetilde{\mathbf{e}}_{c_i}$ $\in \mathbb{R}^{d}$ are the transformed representations.
In this way, redundant correlations in various modalities including texts, images, and cross-modalities could be alleviated, which is helpful to learn universal and robust multi-modal item content representations.

Since there is often large semantic gaps between different domains (\eg ``Apple'' in texts for ``Fruits'' and ``Electronics'' domains),
inspired by UniSRec~\cite{UniSRec2022},
we utilize the mixture-of-expert (MoE) architectures~\cite{MOE} to learn more universal and robust item content representations.
Different from UniSRec, we design the MoEs separately for texts, images and cross-modalities, respectively.
These MoEs can alleviate semantic gaps between different domains and boost the robustness of item content representations.
To be specific, we incorporate $G_1$, $G_2$ and $G_3$ parametric whitening modules as the experts for textual, visual, and cross-modal MoEs, respectively and then construct the MoE projectors based on parameterized routers:
\begin{align}
    &\hat{\mathbf{e}}_{t_i} = \sum_{k=1}^{G_1} g{_{t_k}} \cdot \widetilde{\mathbf{e}}_{t_i}^{(k)},\\
    &\hat{\mathbf{e}}_{p_i} = \sum_{k=1}^{G_2} g{_{p_k}} \cdot \widetilde{\mathbf{e}}_{p_i}^{(k)},\\
    &\hat{\mathbf{e}}_{c_i} = \sum_{k=1}^{G_3} g{_{c_k}} \cdot \widetilde{\mathbf{e}}_{c_i}^{(k)},
\end{align}
where $\widetilde{\mathbf{e}}_{t_i}^{(k)}$, $\widetilde{\mathbf{e}}_{p_i}^{(k)}$ and~$\widetilde{\mathbf{e}}_{c_i}^{(k)}$ are the outputs of the $k$-th parametric whitening modules (Eqn.~\eqref{eq:whitening_1},Eqn.~\eqref{eq:whitening_2} and Eqn.~\eqref{eq:whitening_3}), 
$g{_{t_k}}$, $g{_{p_k}}$ and $g{_{c_k}}$ are the corresponding combination weights from the routing vectors $\bm{g}_t \in \mathbb{R}^{G_1}$, $\bm{g}_p \in \mathbb{R}^{G_2}$ and $\bm{g}_c \in \mathbb{R}^{G_3}$ by the gating routers:
\begin{align}
    \bm{g}_t =& \operatorname{Softmax}\left(\mathbf{e}_{t_i}\cdot \bm{W}_4 + \bm{b}_4 \right),\\
    \bm{g}_p =& \operatorname{Softmax}\left(\mathbf{e}_{p_i}\cdot \bm{W}_5 + \bm{b}_5 \right),\\
    \bm{g}_c =& \operatorname{Softmax}\left(\mathbf{e}_{c_i}\cdot \bm{W}_6 + \bm{b}_6 \right),
    \label{eq:router}
\end{align}
where $\bm{W}_4\in \mathbb{R}^{d_1 \times G_1}$, $\bm{W}_5\in \mathbb{R}^{d_2 \times G_2}$, $\bm{W}_6\in \mathbb{R}^{d_3 \times G_3}$ and $\bm{b}_4\in \mathbb{R}^{G_1}$, $\bm{b}_5\in \mathbb{R}^{G_2}$, $\bm{b}_6\in \mathbb{R}^{G_3}$ are learnable parameters.
Here, we utilize $\mathbf{e}_{t_i}$, $\mathbf{e}_{p_i}$, $\mathbf{e}_{c_i}$ originally from pre-trained models as the inputs of router modules, because they contain obviously domain-specific semantic gaps for recommendation tasks.
Finally, we obtain the textual representation $\hat{\mathbf{e}}_{t_i}$, visual representation $\hat{\mathbf{e}}_{p_i}$, and cross-modal representation $\hat{\mathbf{e}}_{c_i}$ of item $v_i$, respectively.

\subsection{Mixed User Behavior Flow Encoder}
\label{sec:mixedflow}
After the above multi-modal item content constructor and multi-domain item projector, we obtain textual, visual, and cross-modal representations of items from different domains.
Next, we aim to learn the user preferences from all interacted domains for next-item predictions in target domains.
A user usually has common user preference between different domains intuitively.
Hence, for a user, it is natural that her/his interacted items from all domains are mixed in the chronological order and then form a mixed user behavior flow for better understanding her/his common user preference more comprehensively.

Given a user behavior sequence in a target domain, we then mix her/his interacted item from both source and target domains to form the mixed user behavior flow.
Next, we construct the mixed user behavior flow encoder to learn the user preferences from all interacted domains for next-item predictions in the target domain.
Here, we adopt the Transformer architecture~\cite{Attention} consisting of multiple multi-head self-attention layers (denoted by $\operatorname{MHA}(\cdot)$) and point-wise feed-forward networks (denoted by $\operatorname{FFN}(\cdot)$, activated by $\operatorname{ReLU}$) as the mixed user behavior flow encoder.
Formally, we have:
\begin{align}
    \label{eq:TRM1}
    \bm{f}^{0}_j = &[\hat{\mathbf{e}}_{t_i};\hat{\mathbf{e}}_{p_i};\hat{\mathbf{e}}_{c_i}] + \bm{p}_j,\\
    \label{eq:TRM2}
    \bm{F}^\ell = &[\bm{f}^\ell_0;\ldots;\bm{f}^\ell_n],\\
    \label{eq:TRM3}
    \bm{F}^{\ell+1} = &\operatorname{FFN}(\operatorname{MHA}(\bm{F}^\ell)),
\end{align}
where ``[;]'' denotes the concatenation operation, $[\hat{\mathbf{e}}_{t_i};\hat{\mathbf{e}}_{p_i};\hat{\mathbf{e}}_{c_i}]\in \mathbb{R}^{3d}$ represents the concatenated textual, visual, cross-modal representation for item $v_i$,
$\bm{p}_j\in \mathbb{R}^{3d}$ is the position encoding,
$n$ is the length of mixed user behavior flow,
and $\bm{F}^\ell$ represents the concatenated representations at each position in the $\ell$-th layer of total $L$ layers.

In addition, we trained the interacted items in the single source domain, and only leverage multi-modal item content representations in pre-training, the input and update process are the similar with Eqn.~(\ref{eq:TRM1}), Eqn.~(\ref{eq:TRM2}), Eqn.~(\ref{eq:TRM3}), but here $n$ is the length of interacted items in the single domain.
In fine-tuning, we maintain an item embedding matrix $\boldsymbol{M}_{\mathcal{V}} \in \mathbb{R}^{|\mathcal{V}| \times 3d}$ in the target domain to project the high dimensional one-hot representation of each item to the low dimensional dense representation, where $\mathcal{V}$ is the item set in the target domain. Then, we can look up the item embedding to obtain the item embedding  $\mathbf{e}_{v_i}\in \mathbb{R}^{3d}$ of item $v_i$.
The input is $\bm{f}^{0}_j =\mathbf{e}_{v_i}+[\hat{\mathbf{e}}_{t_i};\hat{\mathbf{e}}_{p_i};\hat{\mathbf{e}}_{c_i}] + \bm{p}_j$ for interacted items from only target domain, and the update process is  same as  Eqn.~(\ref{eq:TRM2}), Eqn.~(\ref{eq:TRM3}).

\subsection{Model Training}
\label{sec:training}
To effectively train the above framework, we further introduce two kinds of contrastive learning (CL) tasks in the pre-training stage for bridging the semantic gaps across domains.
By contrasting multi-modal items and sequences from different domains,
we aim to capture their semantic correlations and differences in pre-training.
For this purpose, we design the following cross-domain sequence-item and cross-domain sequence-sequence CL tasks.

\subsubsection{Cross-domain sequence-item CL task}
The cross-domain sequ- ence-item CL task aims to capture the correlations and differences
between user behaviors' contexts and potential next items.
Specifically, we adopt the next items from different instances in the same training batch as negative samples.
Taking a batch of $T$ training instances as an example, 
where each training instance is a pair of the user behaviors' contextual representation and the positive next item,
we encode them into the representations $\{(\bm{u}_{1}, \bm{e}_{1}), \ldots, (\bm{u}_{T}, \bm{e}_{T})\}$, where $\bm{u}$ represents the user behaviors' contextual representation, and $\bm{e}$ denotes the representation of the positive next item.
Then, we formalize the cross-domain sequence-item CL loss as follows:
\begin{equation}
    \mathcal{L}_{CSI} = -\sum_{j=1}^{T} \log \frac{\exp{\left(\bm{u}_j\cdot\bm{e}_j/\tau\right)}}{\sum_{j^{\prime}=1}^{T} \exp{\left(\bm{u}_j\cdot\bm{e}_{j^{\prime}}/\tau\right)}},
\end{equation}
where $\tau>0$ is a temperature parameter, and in-batch next items from each instance are regarded as negative samples.
As the training batches are set up randomly, the in-batch negative samples $\{\bm{e}_{j^{\prime}}\}$ are a set of items from multiple domains.

\subsubsection{Cross-domain sequence-sequence CL task} 
We further adopt a cross-domain sequence-sequence CL task among multi-domain user behavior sequences, in order to discriminate the augmenting user behavior sequences from multi-domain user behavior sequences.
The augmenting user behavior sequences $\hat{\bm{u}}_{j}$ are generated by randomly dropping a proportion of items in the original user behavior sequences $\bm{u}_j$. Formally, we have:
\begin{equation}
    \mathcal{L}_{CSS} = -\sum_{j=1}^{T} \log \frac{\exp{\left(\bm{u}_j\cdot \hat{\bm{u}}_{j}/\tau\right)}}{\sum_{j^{\prime}=1}^{T} \exp{\left(\bm{u}_j\cdot\bm{u}_{j^{\prime}}/\tau\right)}},
\end{equation}
where the augmenting user behavior sequences $\hat{\bm{u}}_{j}$, and other user behavior sequences $\bm{u}_{j^{\prime}}$ in the same batch are regarded as positive and negative samples, respectively.

\subsubsection{Optimization objectives for pre-training}
In pre-training, we jointly optimize $\mathcal{L}_{CSI}$ and $\mathcal{L}_{CSS}$ as follows:
\begin{equation}
    \mathcal{L}_{\text{pre-training}} = \mathcal{L}_{CSI} + \lambda \cdot \mathcal{L}_{CSS},
\end{equation}
where $\lambda>0$ is the weight of cross-domain sequence-sequence CL loss.

\subsubsection{Efficiently and effectively fine-tuning}
Since we aim to design a universal and robust pre-trained MDR framework for better serving real-world multi-domain recommendation scenarios, it is highly essential to fine-tune efficiently and effectively.
Specifically, we only fine-tune a small proportion of parameters from the multi-domain item projector for necessary domain adaptation, and leverage the \emph{mixed user behavior flow encoder} and the \emph{single-domain user behavior sequence encoder} to learn the common and domain-specific user preferences, respectively.
The multi-modal item content representations of mixed user behavior flows are fed into the \emph{mixed user behavior flow encoder}, and the  multi-modal item content representations of user behavior sequence in the target domain incorporating item embeddings are sent into the \emph{single-domain user behavior sequence encoder}.
Then, we fuse the next-item predictions from both mixed and single-domain encoders by equal weights, and optimize the widely used cross-entropy loss to fine-tune parameters of the multi-domain item projector.
And we can observe that the proposed multi-domain item projector can quickly adapt to a new target domain. For the user preference representation $\bm{u}_m$ based on mixed user behavior flows, the next-item prediction probability can be formalized as follows:
\begin{equation}
    \text{P$_m$}(v_{i+1} | \bm{u}_m) = \operatorname{Softmax}\left(\bm{u}_m \cdot [\hat{\mathbf{e}}_{t_{i+1}};\hat{\mathbf{e}}_{p_{i+1}};\hat{\mathbf{e}}_{c_{i+1}}]\right),\label{eq:p1}
\end{equation}
where $v_{i+1}$ is the candidate item, and $\bm{u}_m$ is the  contextual representation ($v_1 \rightarrow v_i$) of mixed user behavior flow.
For the user preference representation $\bm{u}_s$ based on user behavior sequence in the target domain, the next-item prediction probability can be formalized as follows:
\begin{equation}
    \text{P$_s$}(v_{j+1} | \bm{u}_s) = \operatorname{Softmax}\left(\bm{u}_s \cdot ([\hat{\mathbf{e}}_{t_{j+1}};\hat{\mathbf{e}}_{p_{j+1}};\hat{\mathbf{e}}_{c_{j+1}}]+\mathbf{e}_{v_{j+1}})\right),\label{eq:p2}
\end{equation}
where $v_{j+1}$ is the candidate item, $\mathbf{e}_{v_{j+1}}$ is the item embedding of  $v_{j+1}$ and $\bm{u}_s$ is the  contextual representation ($v_1 \rightarrow v_j$) of user behavior sequence in the target domain. Finally, we fuse the next-item predictions in Eqn.~(\ref{eq:p1}) and Eqn.~(\ref{eq:p2}) by equal weights and optimize the cross-entropy loss to fine-tune parameters of the multi-domain item projector.

\subsection{Discussion}
\label{sec:discussion}
We compare the proposed \ourmodel~with the related recommendation models,
in order to highlight the novelty and differences of \ourmodel.
The detailed comparisons of the input types and transfer learning \wrt these approaches are presented in Appendix~\ref{appendix:comparison}.

\textbf{Sequential recommendation approaches} could be roughly divided into two groups, \ie general sequential recommendation (\ie GRU4Rec and SASRec) and pre-training sequential recommendation (\ie S$^3$-Rec and PeterRec).
The first group relys on explicit item IDs to construct the sequential model, 
therefore, these approaches can't perform well under the zero-shot or few-shot item setting with new items.
The second group mainly pre-train their model on sequences by item IDs and textual information, or only item IDs with overlapping users across domains.
However, these methods do not consider other informative modality such as visual or visiolinguistic information,
which is very important in real-world recommendations.
In contrast, our proposed approach can learn more universal and robust item representations by leveraging cross-modality item contents.

\textbf{Multi-domain recommendation approaches} utilize the auxiliary information from source domains to improve the performance in target domains (\eg RecGURU).
However, most of these approaches require overlapping users or items as anchors.
In another line of work, the pre-trained recommendation models are adopted (\eg UniSRec),
but these methods only consider the item texts to be bridges across domains, ignoring other informative modality.
Therefore, we propose the cross-modality item content constructor and robust multi-domain item projectors to tackle the generalization and robustness of existing methods.

\textbf{Multi-modal recommendation approaches} mainly focus on semantic fusion,
or constructing shadow fusion network to achieve semantic alignments (\eg VECF and PMGT).
However, most of these approaches with auxiliary data (even after being aligned) can not naturally fit into the semantic space for recommender systems.
The proposed \ourmodel~adopts the robust multi-domain item projectors, which consists of the parametric whitening and MoE to project the domain-specific modality information of items into a general item space.
The pre-trained contrastive tasks designed for multiple source domains boost acquisitions of the common and domain-specific knowledge from different domains.
In addition, \ourmodel~effectively improves robustness under the real-world recommendation scenarios with missing or noisy item contents.

\section{Experiments}
\subsection{Experimental Setup}

\paratitle{Datasets.}
To evaluate the performance of the proposed approach, we conduct experiments in the pre-trained source and target domain settings.
(1) Pre-trained source domain datasets: we select three categories from Amazon review datasets~\cite{AmazonReview}, ``Home and Kitchen'', ``Clothing, Shoes and Jewelry'', and ``Office Products'' as the source domain datasets for pre-training.
(2) Target domain datasets: we select another five categories from Amazon review datasets~\cite{AmazonReview}, ``Prime Pantry'', ``Musical Instruments'', ``Electronics'', ``Arts, Crafts and Sewing'' and ``Sports and Outdoors'' as the target domain datasets for evaluating the proposed approach.
The statistical details of the pre-processed datasets are shown in Table~\ref{tab:datasets}, including the number of users, items, interactions, average length of interactions per user, and sparsity.

\begin{table}[ht!] %
	\caption{Statistics of the pre-processed datasets. ``Avg. $n$'' denotes the average length of item sequences.
	}
	\label{tab:datasets}
        \centering
	\resizebox{0.99\columnwidth}{!}{
	\begin{tabular}{l *{5}{r}}
		\toprule
		\textbf{Datasets} & \textbf{\#Users} & \textbf{\#Items} & \textbf{\#Inters.} & \textbf{Avg. $n$} & \textbf{Sparsity} \\
		\midrule
		Home   & 731,913 & 185,552 & 6,451,926 & 8.82 & 99.99\% \\
		Clothing & 39,387 & 23,033 &  237,488  & 6.03 & 99.97\%  \\
		Office &  87,436 & 25,986 & 684,837 & 7.84 & 99.97\%  \\
		\midrule
		Pantry &   13,101 & 4,898 & 126,962 & 9.69 & 99.82\% \\
		Instruments  & 24,962 & 22,886 & 208,926 & 8.37 & 99.93\%  \\
		Electronics & 192,403 &  63,001 &  1,689,188 & 8.78 & 99.99\%  \\
	    Arts    & 45,486 & 21,019 & 395,150 & 8.69 & 99.96\% \\
		Sports   & 87,436 & 25,986 & 684,837 & 7.84 &99.95\% \\
		\bottomrule
	\end{tabular}
	}
\end{table}

\paratitle{Evaluation Metrics.}
The proposed approach is evaluated on the next item prediction task and we apply the \textsl{leave-one-out} strategy for evaluation as in previous works~\citep{S3-Rec2020}.
We utilize recall (Recall) and normalized discounted cumulative gain (NDCG) on top-$K$ ranked items to evaluate the performance.
For each user interaction sequence, the last item is used as the test data, the item before the last one is used as the validation data, and the remaining interaction records are used for training.
We rank the ground-truth item of each sequence among all the other items for evaluation on test set, and finally report the average score of all test users.

\noindent
\textbf{Baseline methods.}
We compare the proposed approach with two major groups of baselines.
(1) \emph{ID-based methods}: SASRec~\cite{SASRec}, BERT4- Rec~\cite{BERT4Rec}, RecGURU~\cite{RecGURU2022}, FDSA~\cite{FDSA2019}, S$^3$-Rec~\cite{S3-Rec2020}.
(2) \emph{Modality-based methods}: ZESRec~\cite{ZESRec2021}, UniSRec~\cite{UniSRec2022}, MM-Rec~\cite{MM-Rec2022}.
The details \wrt these baseline methods are presented in Appendix~\ref{appendix:baseline}.

\noindent
\textbf{Implementation Details.}
The proposed \ourmodel~ is implemented based on an open-sourced framework RecBole~\cite{zhao2022recbole} in PyTorch.
In particular, for the fairness of comparisons, we optimize all the methods using Adam optimizer with the batch size of 2048 and carefully search the hyper-parameters of all the compared approaches.
We adopt early stopping with the patience of 10 epochs to prevent overfitting, and NDCG@10 is set as the indicator.
We tune the learning rate in \{0.0003, 0.001, 0.003, 0.01\} and the dimension $d$ of item embedding in \{64, 128\}.
We pre-train the proposed approach for 300 epochs with $G=8$ experts and $\lambda=1e^{-3}$.

\subsection{Overall Performance}

\begin{table*}[!t]
\centering
\caption{Performance comparison of different recommendation models. The best and the second-best performances are denoted in bold and underlined fonts, respectively. ``Improv.'' indicates the relative improvement ratios of \ourmodel~over the best performance baselines. ``*'' indicates that the improvements are significant over all baseline methods ($t$-test with $p<0.05$).}
\label{tab:performance}
\resizebox{2.1\columnwidth}{!}{
\begin{tabular}{@{}c|l|ccccc|cccc|c@{}}
\toprule
\multicolumn{1}{c|}{\multirow{2}{*}{\textbf{Dataset}}} & \multicolumn{1}{c|}{\multirow{2}{*}{\textbf{Metric}}}   & \multicolumn{5}{c|}{\textbf{ID-based}} & \multicolumn{4}{c|}{\textbf{Modality-based}} & \multicolumn{1}{c}{\multirow{2}{*}{\textbf{Improv.}}} \\ \cmidrule(lr){3-7} \cmidrule(lr){8-11} 

  &  & \textbf{SASRec} & \textbf{BERT4Rec} & \textbf{RecGURU} & \textbf{FDSA} & \textbf{S$^3$-Rec} & \textbf{ZESRec} & \textbf{UniSRec} & \textbf{MM-Rec}  & \textbf{\ourmodel} & \\ \midrule

 \multirow{6}{*}{Pantry} & Recall@5 & 0.0271 & 0.0284 & 0.0201 & 0.0230 & 0.0342 & 0.0314   &  0.0378   & \underline{0.0384}    & \textbf{0.0439}* & +14.32\%  \\
  & Recall@10 & 0.0460 & 0.0449 & 0.0279 & 0.0379 & 0.0441 & 0.0527 & \underline{0.0653}  & 0.0641  &  \textbf{0.0741}* & +13.48\% \\
   & Recall@20 & 0.0756 & 0.0768 & 0.0495 & 0.0591 & 0.0712 & 0.0892 & 0.1031  & \underline{0.1050}    & \textbf{0.1163}* & +10.76\% \\
 & NDCG@5 & 0.0142 & 0.0153 & 0.0114 & 0.0152 & 0.0179 & 0.0197 & 0.0219 & \underline{0.0223}   & \textbf{0.0248}* & +11.21\% \\
 & NDCG@10 & 0.0202 & 0.0213 & 0.0152 & 0.0201 & 0.0213 & 0.0249 & 0.0302  & \underline{0.0309}   & \textbf{0.0337}* & \ +9.06\% \\
 & NDCG@20 & 0.0276 & 0.0281 & 0.0185 & 0.0253 & 0.0295 & 0.0335 & 0.0402  & \underline{0.0407}   & \textbf{0.0436}* & \ +7.13\% \\
 \midrule

 \multirow{6}{*}{Instruments} & Recall@5 & 0.0860 & \underline{0.0892} & 0.0612 & 0.0865 & 0.0879 & 0.0842 & 0.0833 & 0.0852 & \textbf{0.0978}* & \ +9.64\% \\
   & Recall@10 & 0.1132 & \underline{0.1201} & 0.1015 & 0.1082 & 0.1107 & 0.1073  & 0.1086  & 0.1103  & \textbf{0.1289}* & \ +7.33\% \\
   & Recall@20 & 0.1483 & \underline{0.1572} & 0.1271 & 0.1374 & 0.1483  & 0.1395 & 0.1450 & 0.1512  & \textbf{0.1697}* & \ +7.95\% \\
   & NDCG@5 & 0.0526 & 0.0553 & 0.0474 & \underline{0.0594} & 0.0581 & 0.0524 & 0.0539 & 0.0559  & \textbf{0.0635}* & \ +6.90\% \\
  & NDCG@10 & 0.0614 & 0.0635 & 0.0536 & \underline{0.0699} & 0.0679  & 0.0596 & 0.0615 & 0.0631  & \textbf{0.0741}* & \ +6.01\% \\
  & NDCG@20 & 0.0703 & 0.0719 & 0.0597 & \underline{0.0811} & 0.0774 & 0.0685 & 0.0712 & 0.0718   & \textbf{0.0842} & \ +3.82\% \\
  \midrule

 \multirow{6}{*}{Electronics} & Recall@5 & 0.0383 & 0.0394 & 0.0299 & \underline{0.0415} & 0.0399 & 0.0413 & 0.0411 & 0.0395 & \textbf{0.0469}* &  +13.01\% \\
  & Recall@10 & 0.0499 & 0.0512 & 0.0384 & \underline{0.0532} & 0.0486 &0.0485& 0.0519 & 0.0503 & \textbf{0.0601}* & +12.96\% \\
   & Recall@20 & 0.0676 & 0.0684 & 0.0491 & \underline{0.0711} & 0.0695 &0.0693& 0.0707 & 0.0675 & \textbf{0.0814}* & +14.49\% \\
 & NDCG@5 & 0.0282 & 0.0273 & 0.0185 & \underline{0.0318} & 0.0295 &0.0278& 0.0305 & 0.0294 & \textbf{0.0351}* & +10.38\% \\
 & NDCG@10 & 0.0319 & 0.0323 & 0.0203 & \underline{0.0349} & 0.0310 &0.0320& 0.0328 & 0.0309 & \textbf{0.0375}* & \ +7.45\% \\
 & NDCG@20   & 0.0364 & 0.0378 & 0.0256 & \underline{0.0403} & 0.0359 &0.0391& 0.0389 & 0.0343 & \textbf{0.0428}* & \ +6.20\% \\

\midrule

\multirow{6}{*}{Arts} & Recall@5 & 0.0804 & 0.0782 & 0.0585 & 0.0797 & \underline{0.0829} & 0.0764 & 0.0791 & 0.0764  & \textbf{0.0917}* & +10.61\%  \\
  & Recall@10 & 0.1083 & 0.1039 & 0.0679 &  0.1021 & \underline{0.1194} & 0.1051 & 0.1095 & 0.1058    &  \textbf{0.1304}* & \ +9.21\% \\
   & Recall@20 & 0.1403 & 0.1354 & 0.0814 & 0.1320 & \underline{0.1492} & 0.1395 & 0.1453  & 0.1397  & \textbf{0.1673}* & \ +12.13\% \\
 & NDCG@5 & 0.0484 & 0.0465 & 0.0326 & \underline{0.0621} & 0.0524 & 0.0473 & 0.0502 & 0.0497  & \textbf{0.0639}* & \ +2.90\% \\
 & NDCG@10 & 0.0575 & 0.0541 & 0.0470 & \underline{0.0693} & 0.0590 & 0.0538 & 0.0599   & 0.0602  & \textbf{0.0715}* & \ +3.17\% \\
 & NDCG@20 & 0.0654 & 0.0625 & 0.0519 & \underline{0.0768} & 0.0667 & 0.0593 & 0.0687   & 0.0681  & \textbf{0.0802}* & \ +4.43\% \\
  \midrule

\multirow{6}{*}{Sports} & Recall@5 & 0.5090 & 0.0521 & 0.0385 & 0.0493 & \underline{0.0514} & 0.0451 & 0.0443  & 0.0419  & \textbf{0.0549}* & \ +6.81\%  \\
   &  Recall@10 & 0.0675 & 0.0693 & 0.0429 & 0.0623 & \underline{0.0697} & 0.0623 & 0.0614  & 0.0595 & \textbf{0.0718}* & \ +3.01\% \\
  & Recall@20 & 0.0910 & 0.0924 & 0.0605 & 0.0821 & \underline{0.0925} & 0.0841 & 0.0864  & 0.0814  & \textbf{0.1043}* & +12.75\%  \\
   & NDCG@5 & 0.0307 & 0.0305 & 0.0218 & \textbf{0.0337} & 0.0304 & 0.0279 & 0.0287 & 0.0304   & \underline{0.0329} & $-$ \\
   & NDCG@10 & 0.0351 & 0.0328 & 0.0257 &  \underline{0.0355} & 0.0365 & 0.0343 & 0.0332  & 0.0339 & \textbf{0.0373}* & \ +5.07\% \\
  & NDCG@20 & 0.0425 & 0.0443 & 0.0324 & \underline{0.0461} & 0.0429 & 0.0405 & 0.0411 & 0.0427  & \textbf{0.0475}* & \ +3.04\% \\

 \bottomrule
 \end{tabular}

}
\end{table*}

We compare the proposed \ourmodel~with the baseline methods on the five target domain datasets.
Note that, we fine-tune the same multi-domain pre-trained model on these source datasets for all downstream domains.
The results of different methods on five datasets are shown in Table~\ref{tab:performance}.

For the ID-based baseline methods,
FDSA and S$^3$-Rec enhanced by textual tags perform better than traditional sequential recommendation methods (\ie SASRec and BERT4Rec) on several datasets,
since item textual tags are used as auxiliary features to improve the performance.
However, S$^3$-Rec sometimes performs worse compared to BERT4Rec. The reason might be that there may be noisy data in the textual tags, and S$^3$-Rec requires the user behavior sequence to be long enough to provide comprehensive contextual information.
The cross-domain recommendation methods (\eg RecGURU) do not perform well, in that these methods become less effective without explicit overlapping users between source and target domains.
As for the modality-based baselines,
these methods have obtained better results compared with ID-based methods.
UniSRec performs better than ZESRec, and a possible reason is that UniSRec incorporates more item texts from source domains to improve the performance in the target domains.
MM-Rec achieves the better performance than UniSRec on several datasets, indicating the effectiveness of modeling multi-modality contents in user behavior sequences.

Finally, the proposed \ourmodel~obtains the best performance in almost all datasets compared with all baselines.
The average improvements compared with the best baseline ranges from +2.90\% to +14.49\% in Recall and NDCG.
Different from these baseline methods, we derive the robust item content representations and the universal mixed user behavior flow encoder from multiple domains via cross-modality item content constructor, robust multi-domain projector and pre-training on multi-domain datasets.



\subsection{Ablation Study}
To evaluate the contribution of each technique or component, we implement several variations of the proposed method:
replacing the cross-modality multi-domain projector in \ourmodel~with a linear layer (\ourmodel $_{w/o~\text{CP}}$),
replacing the textual multi-domain projector with a linear layer (\ourmodel $_{w/o~\text{TP}}$),
replacing the visual multi-domain projector with a linear layer (\ourmodel $_{w/o~\text{VP}}$),
replacing both visual and textual multi-domain projectors with linear layers (\ourmodel$_{w/o~\text{VT}}$),
replacing cross-modality, visual and textual multi-domain projectors with linear layers (\ourmodel$_{w/o~\text{CVT}}$),
replacing the mixed user behavior flow strategy with only considering the target domain user behaviors (\ourmodel$_{w/o~\text{MIX}}$),
removing item IDs in the target domain (\ourmodel$_{w/o~\text{ID}}$),
and removing the pre-trained contrastive learning tasks (\ourmodel$_{w/o~\text{CL}}$).


\begin{table}[!htbp]
\large
\centering
\caption{Ablation study of \ourmodel~ on``Pantry'' and ``Arts''.}
\label{tab:ablation}
\resizebox{0.99\columnwidth}{!}{
\begin{tabular}{l|cc|cc}
\toprule
\multicolumn{1}{c|}{\multirow{2}{*}{Variant}} & \multicolumn{2}{c|}{Pantry} & \multicolumn{2}{c}{Arts}  
\\ \cmidrule(lr){2-3} \cmidrule(lr){4-5}

& Recall@10 & NDCG@10 & Recall@10 & NDCG@10 \\ 

\midrule

\ourmodel & \textbf{0.0741} & \textbf{0.0337} & \textbf{0.1304} & \textbf{0.0715}  \\

\midrule

\ourmodel$_{w/o~\text{CP}}$ & 0.0689 & 0.0316 & 0.1193 & 0.0614  \\
\ourmodel$_{w/o~\text{TP}}$ & 0.0722 & 0.0325 & 0.1249 & 0.0667   \\
\ourmodel$_{w/o~\text{VP}}$ & 0.0731 & 0.0327 & 0.1242 & 0.0673  \\
\ourmodel$_{w/o~\text{VT}}$ & 0.0712 & 0.0320 & 0.1229 & 0.0661 \\
\ourmodel$_{w/o~\text{CVT}}$ & 0.0601 & 0.0295 & 0.1163 & 0.0585  \\
\ourmodel$_{w/o~\text{MIX}}$ & 0.0674 & 0.0309 & 0.1197 & 0.0639  \\
\ourmodel$_{w/o~\text{ID}}$ & 0.0709 & 0.0321 & 0.1255 & 0.0683   \\
\ourmodel$_{w/o~\text{CL}}$ & 0.0619 & 0.0304 & 0.1190 & 0.0603  \\

\bottomrule

\end{tabular}
}
\end{table}

In Table~\ref{tab:ablation} we have: firstly, it would lead to the result decrease when replacing cross-modality, visual or textual item content constructors with linear layers,
indicating the effectiveness of the cross- and mono-modality multi-domain projectors.
Among them, The variant \ourmodel$_{w/o~\text{CP}}$  has poor performance, 
showing that the cross-modality multi-domain projector is the key component to improve the representation capacity.
Secondly, the usage of mix user behavior flow enhances the user preference modeling, thus obtaining a better understanding of the user's global preference.
Finally, the performance degradation of \ourmodel$_{w/o~\text{CL}}$ demonstrates the effectiveness of the pre-trained contrastive learning tasks.
To sum up, all the proposed techniques or components are useful to improve the performance.

\subsection{Robustness Analysis on Modality Noises}
\begin{figure}[!htbp]
    \centering
    \includegraphics[width=0.99\linewidth]{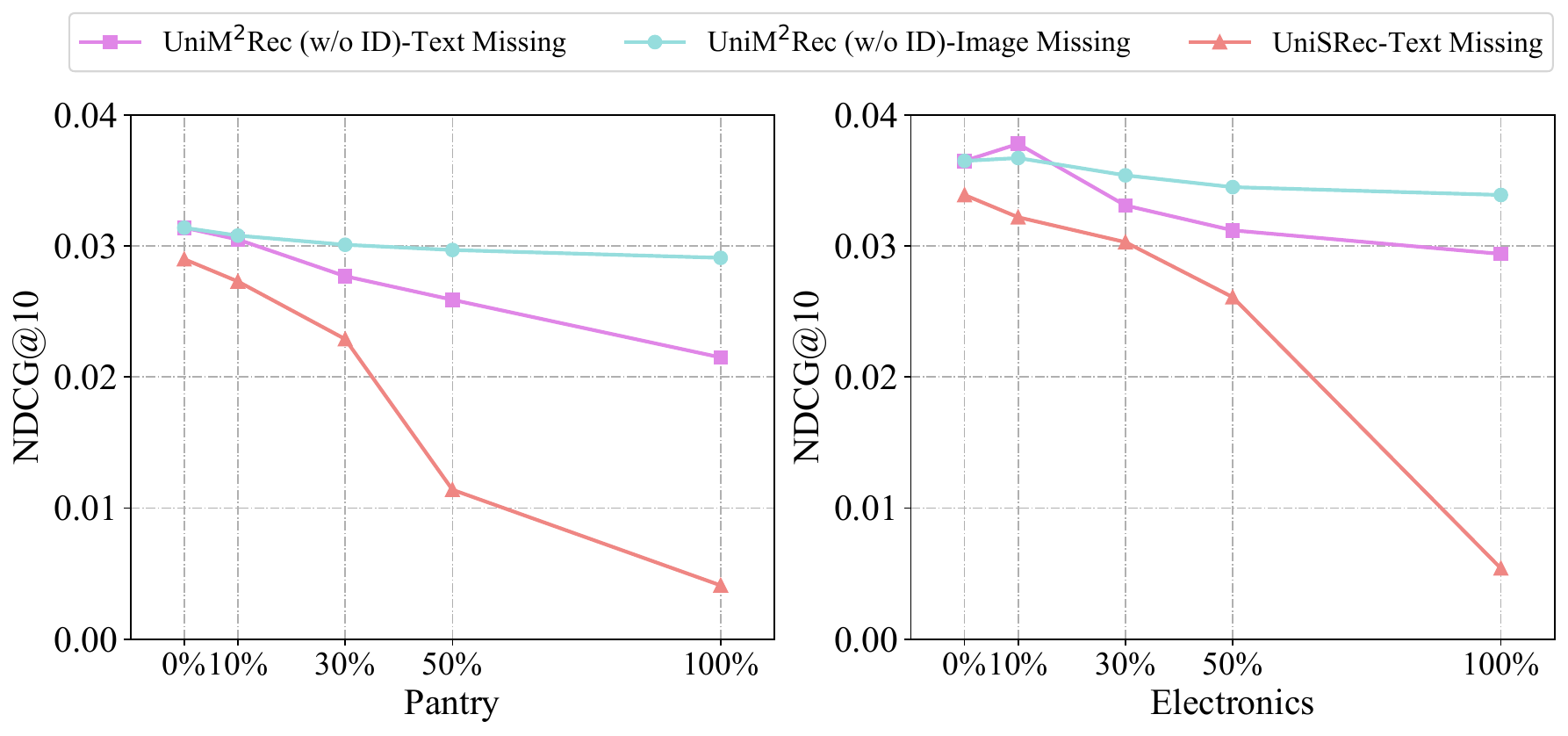}
    \caption{Robustness Analysis on modality noises in target domains (``Pantry'' and ``Arts'').}
    \label{fig:missing}
\end{figure}

We compare \ourmodel$_{w/o~\text{ID}}$ with UniSRec to validate the robustness of our item content representations against possible modality noises.
We randomly remove a certain proportion of item texts or images as the missing contents in the real-world recommendation scenarios, and analyze the trend of NDCG@10 varying with the text or image missing ratios of 10\%, 30\%, 50\% and 100\% respectively.
The experimental results on ``Pantry'', ``Electronics'' and ``Arts'' are shown in Figure~\ref{fig:missing}.
The performance of UniSRec substantially drops with the growing missing ratios of item texts.
In contrast, the performance of \ourmodel$_{w/o~\text{ID}}$ drops slowly as the missing ratios of item texts or images grow.
Therefore, the proposed \ourmodel~possesses the excellent robustness in real-world recommendation scenarios.

\subsection{Analysis on Pre-training Data Sizes}
\begin{figure}[!hbpt]
    \centering
    \includegraphics[width=0.99\linewidth]{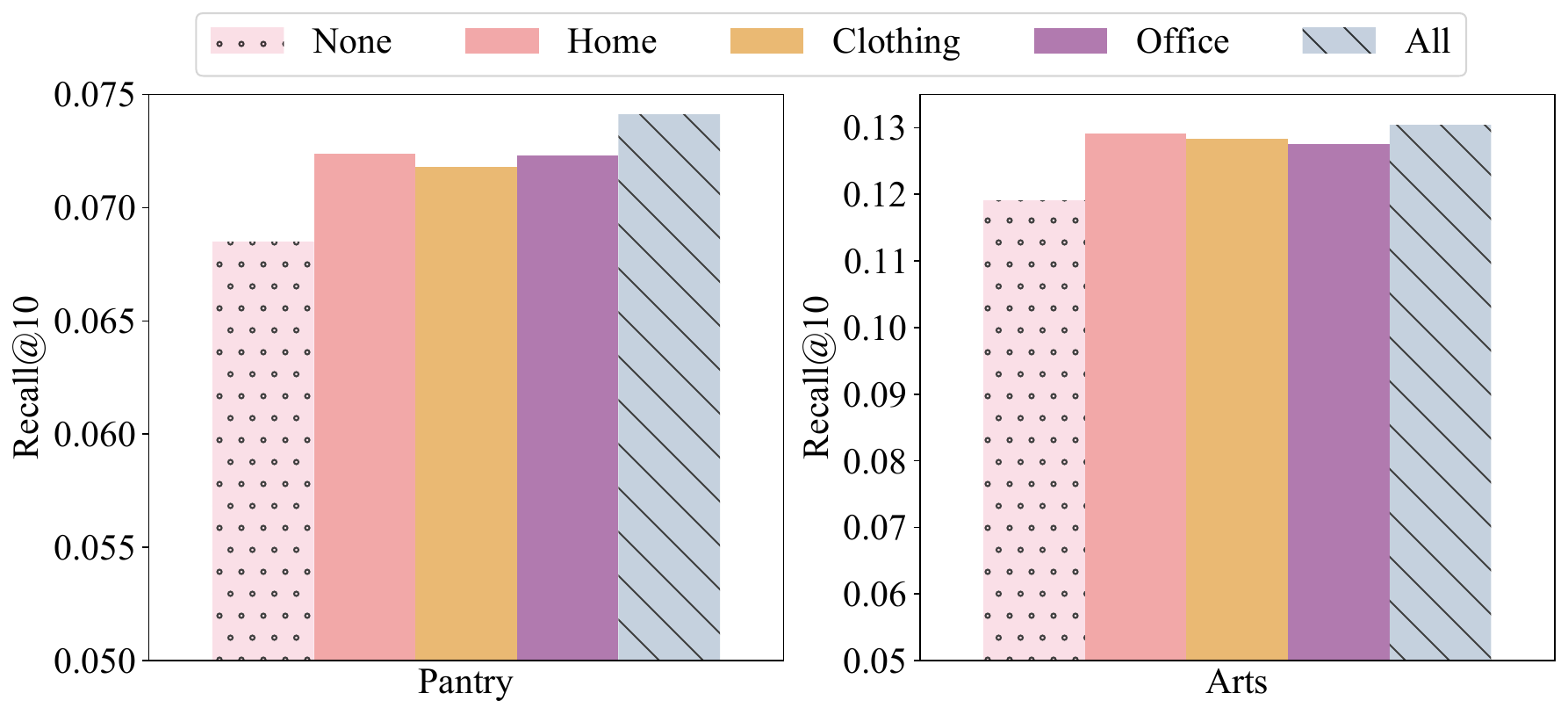}
    \caption{Performance comparisons in target domains (``Pantry'' and ``Arts'') \wrt different pre-trained source domain datasets. ``All'' denotes the model pre-trained on all three datasets, and ``None'' denotes the training from scratch.}
    \label{fig:pre-training}
\end{figure}

We further evaluate \ourmodel~with different amounts of pre-training domains to verify the effectiveness of our pre-training.
The experimental results are reported in Figure~\ref{fig:pre-training}.
We can observe that removing any pre-trained dataset from source domains would lead to the performance degradation, which indicates that adopting the multiple source domain datasets for pre-training can better learn transferable knowledge among different domains and improve the performance in target domains. It is likely that more pre-training datasets will lead to further improvements especially when there are certain (modality) correlations among pre-training and target domains, which will be explored in the future.

\subsection{Analysis on Mixed User Behavior Flow}
Since the multi-domain preferences of a user are helpful to the prediction in target domains intuitively, it is natural that the user's historical behaviors in multiple source domains are directly mixed in the chronological order and fed into the \emph{mixed user behavior flow encoder}.
The experimental results are reported in Table~\ref{tab:mix}.
Note that, \ourmodel$_{w/~\text{TD}}$ denotes leveraging both the mixed user behavior flow and the only target domain behaviors respectively in \ourmodel, and then fusing their predictions.
We can observe that the novel mixed user behavior flow can mostly improve the recommendation performance, whether alone or together with the only target domain behaviors.

\begin{table}[!hbpt]
\large
\centering
\caption{Mixed user behavior flow analysis of \ourmodel~.}
\label{tab:mix}
\resizebox{\columnwidth}{!}{
\begin{tabular}{l|cc|cc}
\toprule
\multicolumn{1}{c|}{\multirow{2}{*}{Variant}} & \multicolumn{2}{c|}{Pantry} & \multicolumn{2}{c}{Arts}   \\ \cmidrule(lr){2-3} \cmidrule(lr){4-5} 

& Recall@10 & NDCG@10 & Recall@10 & NDCG@10   \\ 

\midrule

\ourmodel & 0.0741 & 0.0337 & 0.1304 & 0.0715   \\
\ourmodel$_{w/o~\text{MIX}}$ & 0.0674 & 0.0309 & 0.1197 & 0.0639  \\
\ourmodel$_{w/~\text{TD}}$ & 0.0749 &0.0331 & 0.1295 & 0.0708 \\

\bottomrule

\end{tabular}
}
\end{table}

\subsection{Analysis on Few-shot Items}
\begin{figure}[!hbpt]
	{
		\begin{minipage}[t]{0.49\linewidth}
			\centering
			\includegraphics[width=1\textwidth]{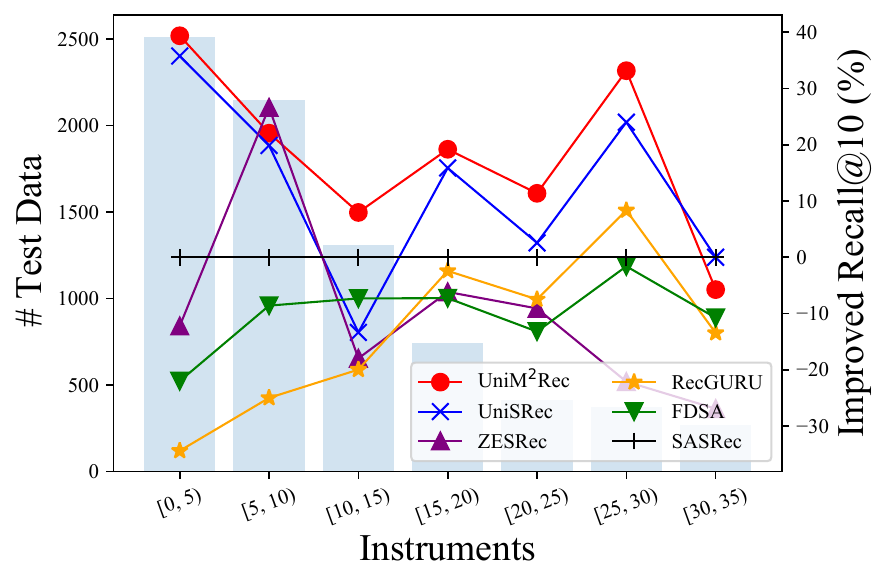}
		\end{minipage}
		\begin{minipage}[t]{0.49\linewidth}
			\centering
			\includegraphics[width=1\textwidth]{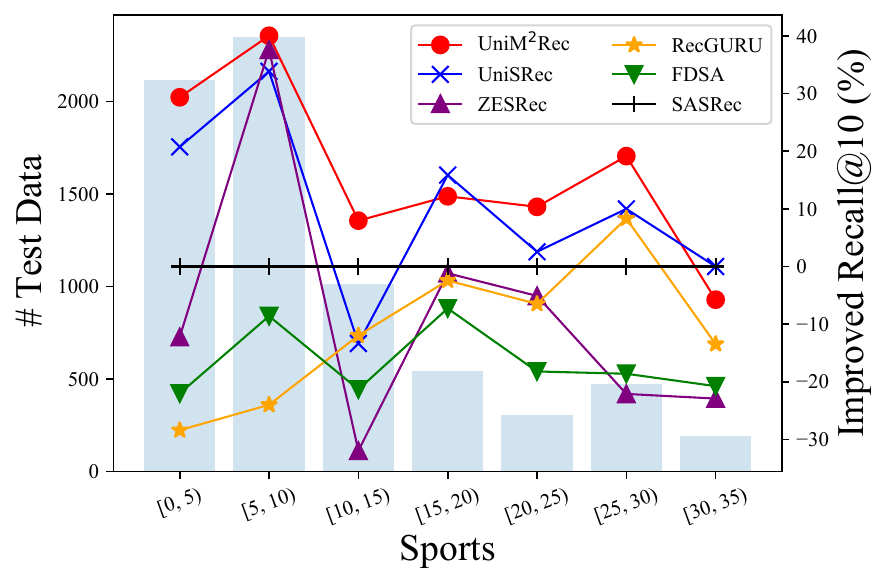}
		\end{minipage}
	}
	\caption{Performance comparison \wrt few-shot items. The bar graph denotes the number of interactions in test data for each group. The line chart denotes the relative improvement ratios on Recall@10 compared with SASRec.}
	\label{fig:few-shot}
\end{figure}

We verify the universal and robust latent representations of cross-modality item contents can alleviate the issue of few-shot item recommendations.
We split the test data into different groups according to the popularity of ground-truth items in the training data,
and then compare the improved ratios of Recall@10 score with the baseline SASRec in each group.
From Figure~\ref{fig:few-shot}, we can observe that the recommenders considering item contents achieve better results in few-shot items. Among them, the proposed \ourmodel~ consistently outperforms other baseline models on both datasets in most cases, especially when the ground truth item is not popular. The results show that few-shot items can benefit more from the cross-modality item content representations in our pre-trained recommendation model.

\section{Conclusion and Future Work}
In this paper,
we propose a novel \ourmodel~to effectively and robustly deal with multi-modal inputs of multi-domain items.
Different from existing MDR methods that rely on item texts to be the bridge across domains, 
the proposed method leverages cross-modality item contents to learn the universal and robust item representations.
With the pre-trained multi-domain recommender, \ourmodel~could be efficiently transferred to new target domains.
Extensive experiments on five real-world datasets demonstrate the effectiveness of the proposed method.

As future work, we will explore more kinds of item contents (\eg videos and audios) to learn more robust item content representations and model more universal user preferences. We will also explore the potential promising usage of large-scale multi-modal models in recommendation.

\bibliographystyle{ACM-Reference-Format}
\bibliography{reference}

\clearpage
\appendix

\section{Comparisons of the input types and transfer learning}
\label{appendix:comparison}

We compare the proposed \ourmodel~with the related recommendation models,
in order to highlight the novelty and differences of \ourmodel.
The detailed comparisons of the input types and transfer learning \wrt these approaches are presented in Table~\ref{tab:comparison}.
We aim to build a universal multi-modal multi-domain pre-trained recommender, which could smoothly deal with the multi-modal item contents and learn the multi-domain preferences, and then can be effectively transferred to new target domains.
Therefore, we compare the proposed \ourmodel~with existing related methods from two aspects (\ie the input types and the transfer learning capability).

For the input types, the proposed \ourmodel~adopts the texts and images in the pre-training stage. Then, it leverages the visual and textual representations of items from pre-training, and item IDs in the serving (\ie fine-tuning) stage.
Table~\ref{tab:comparison} shows the input types that the existing related methods can deal with in detail.
Hence, the proposed \ourmodel~is a universal recommender that can address the various different types of item contents (\eg texts and images) and item IDs in the pre-training/fine-tuning stages.

For the transfer learning capability, the proposed \ourmodel~pre-trains multi-domain datasets in the source domains in order to obtain the universal transfer learning capability. Compared with the conventional cross-domain recommendation methods (\eg CCDR), the proposed \ourmodel~does not require the overlapping users or items, which is suitable for more recommendation scenarios.
Compared with the pre-trained recommenders such as S$^3$-Rec and P5, the proposed \ourmodel~can pre-train the multi-domain datasets, which can help to improve the model generalizations, and be effectively transferred to the new target domains.
Compared with another line of  pre-trained recommenders such UniSRec and Recformer, the proposed \ourmodel~can smoothly deal with multi-modal item contents (\eg images and texts), and fully consider the mixed user behavior flow from the source domains, which can better address the  data sparsity issue in recommender systems.
Compared with the multi-modal recommenders such as MM-Rec, the proposed \ourmodel~adopts the parameter whitening and MoE to learn the multi-modal item representations, which can better adapt to new domains.
Therefore, the proposed \ourmodel~is a universal and robust recommender that can be effectively transferred to new target domains and better address the data sparsity issue via the mixed user behavior flow in recommender systems.

\begin{table}[!thbp]

\small
\centering
\caption{Comparisons of the input types and transfer learning \wrt several representative approaches.}

\resizebox{1\columnwidth}{!}{
\begin{tabular}{@{}lccccccc@{}}
\toprule
\multicolumn{1}{c}{\multirow{2}{*}{Method}} & \multicolumn{3}{c}{Input Types} & \multicolumn{4}{c}{Transfer Learning} \\ \cmidrule(l){2-4} \cmidrule(l){5-8}
          & ID & Text & Image & Pre-training & Adaption & Mixed & Cross-domain   \\

\midrule

SASRec~\cite{SASRec} & \textcolor{teal}{\CheckmarkBold}& \textcolor{purple}{\XSolidBrush}& \textcolor{purple}{\XSolidBrush}& \textcolor{purple}{\XSolidBrush}& \textcolor{purple}{\XSolidBrush}& \textcolor{purple}{\XSolidBrush}& \textcolor{purple}{\XSolidBrush} \\

BERT4Rec~\cite{BERT4Rec} & \textcolor{teal}{\CheckmarkBold} & \textcolor{purple}{\XSolidBrush} & \textcolor{purple}{\XSolidBrush} & \textcolor{purple}{\XSolidBrush} & \textcolor{purple}{\XSolidBrush} & \textcolor{purple}{\XSolidBrush} & \textcolor{purple}{\XSolidBrush} \\

RecGURU~\cite{RecGURU2022}  & \textcolor{teal}{\CheckmarkBold} & \textcolor{purple}{\XSolidBrush} & \textcolor{purple}{\XSolidBrush} & \textcolor{teal}{\CheckmarkBold} & \textcolor{purple}{\XSolidBrush} & \textcolor{purple}{\XSolidBrush} & \textcolor{teal}{\CheckmarkBold} \\

FDSA~\cite{FDSA2019} & \textcolor{teal}{\CheckmarkBold} & \textcolor{teal}{\CheckmarkBold} & \textcolor{purple}{\XSolidBrush} & \textcolor{purple}{\XSolidBrush} & \textcolor{purple}{\XSolidBrush} & \textcolor{purple}{\XSolidBrush} & \textcolor{purple}{\XSolidBrush}  \\

S$^3$-Rec~\cite{S3-Rec2020}  & \textcolor{teal}{\CheckmarkBold} & \textcolor{teal}{\CheckmarkBold} & \textcolor{purple}{\XSolidBrush} & \textcolor{teal}{\CheckmarkBold} & \textcolor{purple}{\XSolidBrush} & \textcolor{purple}{\XSolidBrush} & \textcolor{purple}{\XSolidBrush} \\ 

CCDR~\cite{CCDR2022} & \textcolor{teal}{\CheckmarkBold} & \textcolor{teal}{\CheckmarkBold} & \textcolor{purple}{\XSolidBrush} & \textcolor{purple}{\XSolidBrush} & \textcolor{purple}{\XSolidBrush} & \textcolor{purple}{\XSolidBrush} & \textcolor{teal}{\CheckmarkBold} \\

P$5$~\cite{P5-2022} & \textcolor{teal}{\CheckmarkBold} & \textcolor{purple}{\XSolidBrush} & \textcolor{purple}{\XSolidBrush} & \textcolor{teal}{\CheckmarkBold} & \textcolor{purple}{\XSolidBrush} & \textcolor{purple}{\XSolidBrush} & \textcolor{purple}{\XSolidBrush} \\

PeterRec~\cite{PeterRec2020} & \textcolor{teal}{\CheckmarkBold} & \textcolor{purple}{\XSolidBrush} & \textcolor{purple}{\XSolidBrush} & \textcolor{teal}{\CheckmarkBold} & \textcolor{purple}{\XSolidBrush} & \textcolor{purple}{\XSolidBrush} & \textcolor{teal}{\CheckmarkBold} \\

ZESRec~\cite{ZESRec2021} & \textcolor{purple}{\XSolidBrush} & \textcolor{teal}{\CheckmarkBold} & \textcolor{purple}{\XSolidBrush} & \textcolor{teal}{\CheckmarkBold}  & \textcolor{purple}{\XSolidBrush} & \textcolor{purple}{\XSolidBrush} & \textcolor{teal}{\CheckmarkBold} \\ 

UniSRec~\cite{UniSRec2022} & \textcolor{purple}{\XSolidBrush} & \textcolor{teal}{\CheckmarkBold} & \textcolor{purple}{\XSolidBrush} & \textcolor{teal}{\CheckmarkBold} & \textcolor{teal}{\CheckmarkBold} & \textcolor{purple}{\XSolidBrush} & \textcolor{teal}{\CheckmarkBold}  \\

MM-Rec~\cite{MM-Rec2022} & \textcolor{purple}{\XSolidBrush} & \textcolor{teal}{\CheckmarkBold} & \textcolor{teal}{\CheckmarkBold} &  \textcolor{teal}{\CheckmarkBold} & \textcolor{purple}{\XSolidBrush} & \textcolor{purple}{\XSolidBrush} & \textcolor{purple}{\XSolidBrush}  \\

Recformer~\cite{Recformer2023} & \textcolor{purple}{\XSolidBrush} & \textcolor{teal}{\CheckmarkBold} & \textcolor{purple}{\XSolidBrush} & \textcolor{teal}{\CheckmarkBold} & \textcolor{teal}{\CheckmarkBold} & \textcolor{purple}{\XSolidBrush} & \textcolor{teal}{\CheckmarkBold} \\

\ourmodel~(Ours) & \textcolor{teal}{\CheckmarkBold} & \textcolor{teal}{\CheckmarkBold} & \textcolor{teal}{\CheckmarkBold} & \textcolor{teal}{\CheckmarkBold} & \textcolor{teal}{\CheckmarkBold} & \textcolor{teal}{\CheckmarkBold} & \textcolor{teal}{\CheckmarkBold}  \\
\bottomrule
\end{tabular}
}
\label{tab:comparison}
\end{table}

\section{Baseline Methods in Detail}
\label{appendix:baseline}

(1) \textbf{ID-based methods}: SASRec, BERT4Rec, RecGURU, FDSA, S$^3$-Rec.
\begin{itemize}
    \item SASRec~\citep{SASRec} is a next-item sequential recommendation method based on the Transformer architecture employing multi-head self-attention mechanism to explore implicit user interactions.
    \item BERT4Rec~\citep{BERT4Rec} is an improvement of SASRec, which contains an additional Cloze objective and bidirectional self-attention structure.
    \item RecGURU~\citep{RecGURU2022} proposes to pre-train user representations via auto-encoder in an adversarial learning paradigm.
    \item S$^3$-Rec~\citep{S3-Rec2020} pre-trains sequential models to predict the correlation between an item and its attributes via mutual information maximization objectives for feature fusion.
    \item FDSA~\citep{FDSA2019} proposes to utilize two self-attentive networks to capture item and feature transition patterns.
\end{itemize}

\noindent
(2) \textbf{Modality-based methods}: ZESRec, UniSRec, MM-Rec.
\begin{itemize}
    \item ZESRec~\citep{ZESRec2021} encodes item text via pre-trained language model as item representations.
    \item UniSRec~\citep{UniSRec2022} is an improvement of ZESRec, which pre-trains item texts as item representations with consideration of multiple domains.
    \item MM-Rec~\citep{MM-Rec2022} proposes a multi-modal network to capture their inherent multi-modal correlations for news recommendation.
\end{itemize}

\noindent
(3) \textbf{Other methods} (without comparisons and the reasons): P5, M6-Rec, VIP5, Recformer, CCDR.

\begin{itemize}
    \item P5~\cite{P5-2022} formulates the multiple recommendation tasks with instruction-based prompts and demonstrates competitive performance on multiple tasks. However, it can not incorporate the multi-domain datasets since it uses item IDs but not modality-based item semantics to to represent items.
    \item M6-Rec~\cite{cui2022m6} proposes to represent the user behaviors as plain texts, and leverage item texts instead of item IDs. However, it is not very universal since the length of text inputs is limited by the language models, similarly, the number of candidate items is also limited.
    \item VIP5~\cite{geng2023vip5} is an improvement of P5, which leverages item IDs and also incorporates the images of items. Therefore, it can not incorporate the multi-domain datasets.
    \item Recformer~\cite{Recformer2023} proposes to formulate items as texts, pre-train the text sequence and then retrieve the next item represented by texts. However, it can not incorporate other informative modalities (\eg images). 
    \item CCDR~\cite{CCDR2022} proposes two intra-domain and inter-domain contrastive learning tasks to enhance cross-domain recommendation, where the aligned users, tags, words, and categories are functioned as the semantic bridges across different domains. However, it relies on overlapped users or items, which is different from our experimental settings.
\end{itemize}

\end{document}